\documentclass[aps,prl,twocolumn,floatfix,showpacs,superscriptaddress]{revtex4}
\usepackage[dvips]{graphicx}
\usepackage{amssymb}
\usepackage{amsmath}

\begin{document}
\title{Light emission patterns from stadium-shaped semiconductor
  microcavity lasers}
\newcommand{\affiliationATR}{\affiliation{Department of Nonlinear
    Science, ATR Wave Engineering Laboratories, 2-2-2 Hikaridai,
    Seika-cho, Soraku-gun, Kyoto 619-0288, Japan}}

\newcommand{\affiliationOPU}{\affiliation{Department of Communication
    Engineering, Okayama Prefectural University, 111 Kuboki, Soja, 
    Okayama 719-1197, Japan}}

\author{Susumu Shinohara}
\affiliationATR

\author{Takehiro Fukushima}
\affiliationATR
\affiliationOPU

\author{Takahisa Harayama}
\affiliationATR

\begin{abstract}
We study light emission patterns from stadium-shaped semiconductor
(GaAs) microcavity lasers theoretically and experimentally. Performing
systematic wave calculations for passive cavity modes, we demonstrate
that the averaging by low-loss modes, such as those realized in
multi-mode lasing, generates an emission pattern in good agreement
with the ray model's prediction. In addition, we show that the
dependence of experimental far-field emission patterns on the aspect
ratio of the stadium cavity is well reproduced by the ray model.
\end{abstract}


\maketitle

To theoretically investigate emission patterns from two-dimensional
microcavity lasers, two approaches have been successfully employed,
one based on the wave description and the other based on the ray
description.
The ray description is regarded as an approximation of the wave
description and can be justified in the short-wavelength limit by the
Eikonal theory only in limited cases, i.e., when ray dynamics becomes
integrable.
It has been a fundamental problem in the field of quantum chaos to
study how the nonintegrable or chaotic ray dynamics manifests itself
in the wave description \cite{QC}.
In spite of the lack of full justification, the ray description has
been practically used for cavities that exhibit chaotic ray dynamics,
and shown useful for giving a simple explanation for the appearance of
emission directionality \cite{Nockel96, Nockel97, Hentschel01,
Schwefel04, Shinohara06, Lebental07, SB.Lee07, JB.Shim06}.

Recent studies on microcavities with low refractive indices
(e.g. $n=1.5$ for polymers and $n=1.36$ for dye-doped liquid jets)
have revealed a remarkable ray-wave correspondence in low-loss cavity
modes; the far-field emission pattern of an individual low-loss cavity
mode exhibits close agreement with the result generated by a ray model
\cite{Schwefel04, Shinohara06, Lebental07, SB.Lee07, JB.Shim06}.
So far, it is not clear whether this ray-wave correspondence is due to
the low refractive indices (i.e., high openness) or a more robust
property holding also for higher refractive index cases.
The present work examines the ray-wave correspondence for
semiconductor (GaAs) cavities with the stadium shape
\cite{Bunimovich}, whose refractive index is much higher ($n=3.3$)
compared to polymers and dye-doped liquid jets.
While there exist several experimental and theoretical works for GaAs
microcavities \cite{Gmachl98, Fukushima04, Tanaka07, Fang07},
systematic analysis has not yet been performed for low-loss cavity
modes in the short-wavelength regime.

In this paper, we show that the far-field emission pattern of an
individual low-loss mode does not always exhibit good agreement with
the ray model's prediction, being different from the cases for the low
refractive index cases.
Nevertheless, a new correspondence is found in the phase space
distributions describing near-field emission patterns.
We demonstrate that close correspondence between the ray and the wave
description in the far field is retrieved by performing the averaging
by low-loss cavity modes.
In addition, we show that an experimental far-field emission pattern,
which can be approximated by the averaged result of low-loss modes,
systematically agrees with the result of the ray model.

\begin{figure}[t]
\includegraphics[width=75mm]{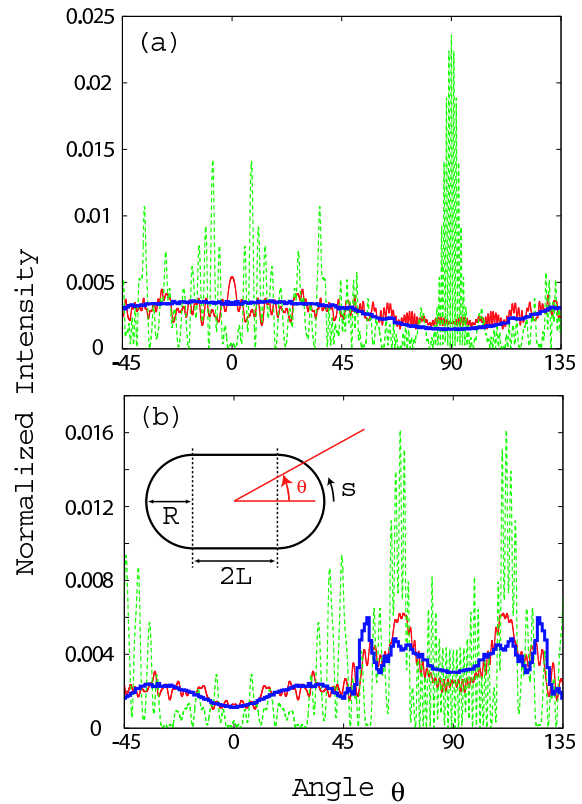}
\caption{(Color) Far-field emission patterns numerically calculated
  for $n=3.3$ stadium cavities with (a) $\epsilon=1.0$ and (b)
  $\epsilon=0.3$. $\theta$ is the far-field angle defined in the inset
  of Fig. \ref{fig:ffp} (b). The blue curves are the patterns obtained
  from the ray model. The green curves are the patterns for cavity
  modes with low loss. The mode in (a) has the seventh lowest loss and
  the mode in (b) has the third lowest loss in the searched $kR$
  range. The red curves are the average of the far-field patterns of
  the 30 lowest-loss modes. We note that all the far-field emission
  patterns are normalized so that the integration becomes unity.}
\label{fig:ffp}
\end{figure}

We define the shape of a stadium cavity in the inset of
Fig. \ref{fig:ffp} (b).
We introduce an aspect ratio parameter $\epsilon=L/R$, where $L$ is
the half length of the linear part and $R$ the radius of the circular
part.
We assume that the refractive index inside the cavity is $n=3.3$ and
$n=1.0$ outside the cavity.
For $\epsilon=1.0$, far-field emission patterns have been obtained
experimentally and their trends have been explained by the ray model
\cite{Fukushima04}.
Whereas the far-field emission pattern is less directional for
$\epsilon=1.0$, by decreasing the value of $\epsilon$ to less than
around 0.3, one finds drastic changes in the far-field emission
patterns.
This drastic change can be associated with the phase space flow
governed by the unstable manifolds of three-bounce periodic orbits,
for which all the bounces occur at the circular parts.
The details of this drastic change will be reported elsewhere.

In Fig. \ref{fig:ffp}, we plot far-field emission patterns for
$\epsilon=1.0$ and $\epsilon=0.3$ generated by a ray model.
In the ray model, we incorporate Fresnel's law for transverse magnetic
(TM) polarization to describe the light leakage at the cavity boundary
\cite{Shinohara06, JW.Ryu06, SY.Lee04, Shinohara07}.
Another method to theoretically obtain far-field emission patterns is
based on wave calculation.
Passive cavity modes are calculated by using, for instance, the
boundary element method \cite{Wiersig03}.
These are the eigensolutions of the Helmholtz equation
$[\nabla^2+n^2k^2]\psi=0$,
where $k$ is a wave number outside the cavity and becomes complex by
imposing the outgoing wave condition at infinity.
For computational simplicity, we assume that the electric field is TM;
we regard $\psi$ as the $z$ component of the electric field and assume
that $\psi$ and its normal derivative $\partial \psi$ are continuous
at the cavity boundary.
%
%
Because of computational limitation, we set the size parameter as
$kR\approx 100$ (i.e., $nkR\approx 330$), which is about half of a
real cavity size used in experiments discussed below.
Since we have previously confirmed the convergence of wave
calculations even when $nkR$ is less than 70 \cite{Shinohara06}, we
think $nkR\approx 330$ is large enough to discuss semiclassical
properties and thus allows one to compare wave calculations with
experiments.
Also, we note $nkR\approx 330$ is much larger compared with recent
previous works \cite{Lebental07,JB.Shim06,Shinohara07}.
For $\epsilon=1.0$ and $\epsilon=0.3$, we obtained in each case around
100 passive cavity modes for $99.95\leq \mbox{Re}\,kR\leq 100.05$.
We are interested in the emission patterns of low-loss modes, since it
is these modes that contribute to lasing.
We plot in Fig. \ref{fig:ffp}, far-field emission patterns for
low-loss modes.
For $n=1.5$, it was relatively easy to confirm the correspondence
between the results of wave calculations and those of the ray model,
since the ray model predicts the appearance of four distinct narrow
peaks in the far-field emission pattern \cite{Shinohara06,Lebental07}.
However, for the less directional cases shown in Figs. \ref{fig:ffp}
(a) and \ref{fig:ffp} (b), the correspondence between wave
calculations and ray simulations becomes less clear.
In particular, in Fig. \ref{fig:ffp} (a), we see a large discrepancy
at $\theta=90^{\circ}$, where $\theta$ is the far-field angle defined
in the inset of Fig. \ref{fig:ffp} (b).
We note that the peaking at $\theta=90^{\circ}$ is not a common
feature of low-loss modes.
By averaging the far-field emission patterns over low-loss modes, we
see that the averaged patterns correspond closely with the ray model's
results.

\begin{figure}[t]
\includegraphics[width=72mm]{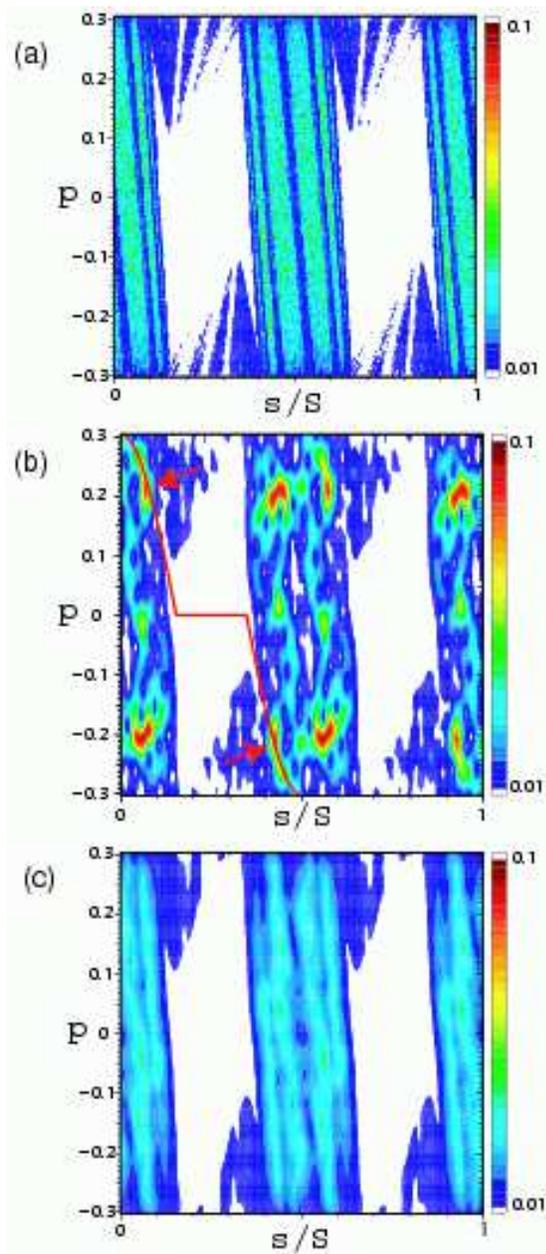}
\caption{(Color) Phase space distributions for a stadium cavity with
  $n=3.3$ and $\epsilon=1.0$. (a) $P(s,p)$ generated by the ray
  model. (b) $H(s,p)$ for a low-loss cavity mode whose far-field
  emission pattern is plotted in Fig. \ref{fig:ffp} (a). The points on
  the red curve contribute to the far-field emission at
  $\theta=90^{\circ}$. High-intensity regions resulting in strong
  far-field emission at $\theta=90^{\circ}$ are indicated by
  arrows. (c) The average of $H(s,p)$ of the 30 lowest-loss modes.}
\label{fig:dist}
\end{figure}

The correspondence between the ray model and the wave description can
be further studied by looking at the near fields.
Here we introduce functions describing near-field emission patterns,
defined in the phase space spanned by the Birkhoff coordinates
$(s,p)$, where $s$ is the curvilinear coordinate along the cavity
boundary and $p$ the momentum component tangential to the cavity
boundary.

In the ray model for stadium cavities, the light intensity inside the
cavity ${\cal E}(t)$ can be asymptotically described as ${\cal
E}(t)\propto \exp[-\gamma_R t]$ \cite{JW.Ryu06}, where the decay rate
$\gamma_R$ can be expressed as \cite{SY.Lee04, Shinohara07}
\begin{equation}
\gamma_R=\int_{0}^{S} ds \int_{-1/n}^{1/n} dp\,P(s,p),
\end{equation}
where $S$ is the total boundary length and we assume momentum is
normalized to unity.
The function $P(s,p)$ describes how much light is transmitted outside
the cavity at a boundary point $s$ in the direction determined by $p$.
We plot $P(s,p)$ for a stadium cavity with $\epsilon=1.0$ in
Fig. \ref{fig:dist} (a).

For a passive cavity mode, on the other hand, the corresponding decay
rate can be expressed as \cite{Shinohara07}
\begin{equation}
\gamma_W=\int_{0}^{S} ds \int_{-1/n}^{1/n} dp\,H(s,p).
\end{equation}
Here, $H(s,p)$ is a Husimi-like phase space distribution calculated
from the wave function at the cavity boundary $\psi(s)$ and its normal
derivative $\partial\psi(s)$, defined by
\begin{equation}
H(s,p)=\mbox{Im}[h_{\psi}^*(s,p)h_{\partial\psi}(s,p)],
\end{equation}
where $h_f(s,p)=\int ds' G^*(s';s,p)f(s')$ and $G(s';s,p)$ is a
coherent state for a one-dimensional periodic system
\begin{equation}
G(s';s,p)=\frac{1}{\sqrt{\sigma\sqrt{\pi}}}\sum_{m=-\infty}^{\infty}
e^{\left\{
-\frac{(s'-s-m S)^2}{2\sigma^2}+ip(s'-s-m S)
\right\}}
\end{equation}
with $\sigma=\sqrt{S/[2n\mbox{Re}(kR)]}$.
Below, we will investigate the correspondence between $H(s,p)$ and
$P(s,p)$.

In Fig. \ref{fig:dist} (b), we plot $H(s,p)$ for a low-loss cavity
mode whose far-field emission pattern is shown in Fig. \ref{fig:ffp}
(a).
We superimpose a set of points [red curve in Fig. \ref{fig:dist} (b)]
giving the far-field emission at $\theta=90^{\circ}$.
One can see that the strong near-field emissions indicated by arrows
result in the strong far-field emission at $\theta=90^{\circ}$, which
can be observed in Fig. \ref{fig:ffp} (a).

\begin{figure}[b]
\includegraphics[width=82mm]{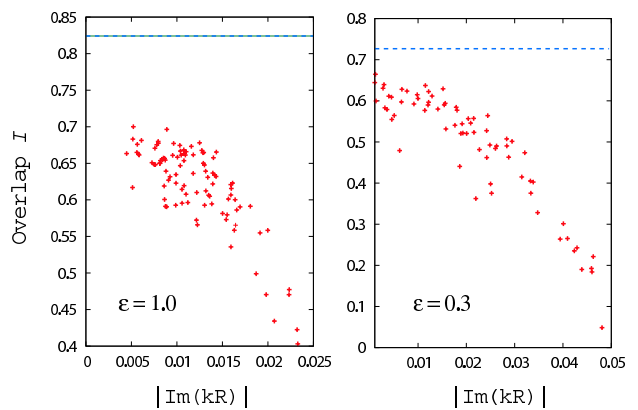}
\caption{(Color online) The overlap $I$ of $\Omega_P$ and $\Omega_H$
  as a function of the loss rate $|\mbox{Im}(kR)|$ for $\epsilon=1.0$
  (left) and $\epsilon=0.3$ (right). The overlap of the high-intensity
  regions of the averaged $H$ and $\Omega_P$ is indicated by a dashed
  line.}
\label{fig:overlap}
\end{figure}

\begin{figure}[h]
\includegraphics[width=72mm]{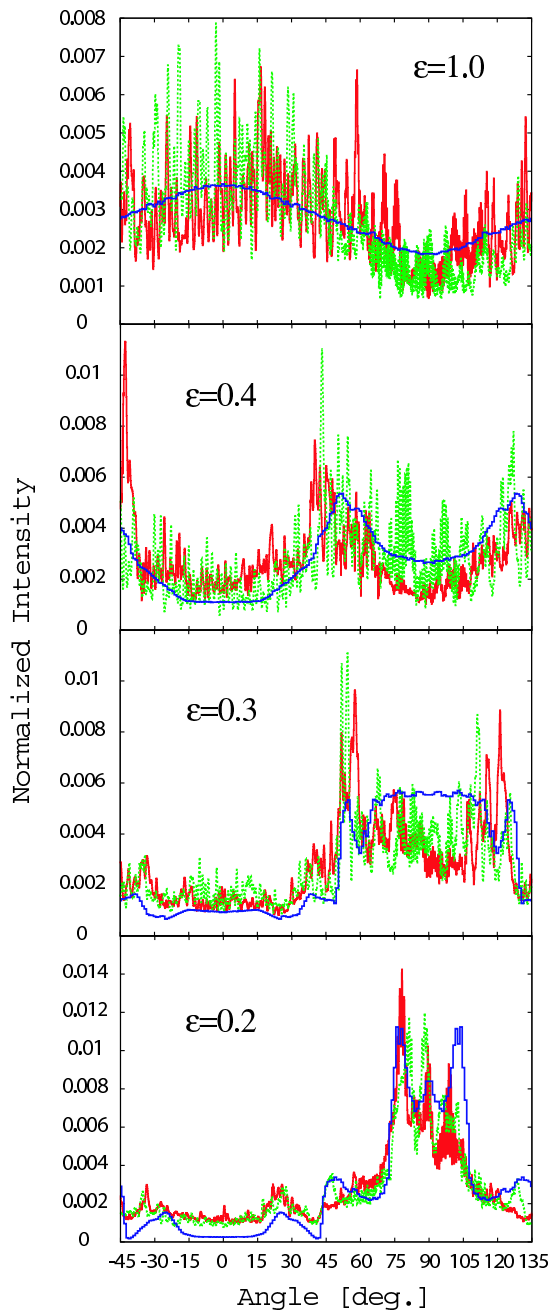}
\caption{(Color) Far-field emission patterns for stadium-shaped
  semiconductor microcavity lasers. Red and green curves are
  experimental data for two different samples fabricated in the same
  manner. Blue curves are data generated by the ray model. In all
  data, intensities are normalized so that the integration becomes
  unity.}
\label{fig:ffp2}
\end{figure}

In Fig. \ref{fig:dist} (b), we can also see that $H(s,p)$ is mainly
supported on the high-intensity regions of $P(s,p)$.
This property has already been found for stadium cavities with $n=1.5$
as a common feature of low-loss modes \cite{Shinohara07}.
Below, we systematically check this property for $n=3.3$ by
introducing a quantity measuring the overlap of the high-intensity
regions of $P(s,p)$ and those of $H(s,p)$.
We define the high-intensity regions of a function $f(s,p)$ as
\begin{equation}
\Omega_{f}=\left\{
(s,p)\,|\,f(s,p)>\bar{f}
\right\},
\end{equation}
where $\bar{f}$ is the average of $f(s,p)$, i.e., 
$\bar{f}=\int_{0}^{S} ds \int_{-1/n}^{1/n} dp f(s,p)/(2S/n)$.
The overlap between $\Omega_P$ and $\Omega_H$ can be defined as
\begin{equation}
I=\frac{\mu(\Omega_P \cap \Omega_H)}{\sqrt{\mu(\Omega_P)\mu(\Omega_H)}},
\end{equation}
where $\mu(\Omega)$ is the Lebesgue measure of the set $\Omega$.
The overlap $I$ takes the maximum value $I=1$ when $\Omega_P=\Omega_H$
and takes the minimum value $I=0$ when $\Omega_P\cap\Omega_H=\phi$.
For instance, the $I$ value for $H(s,p)$ shown in Fig. \ref{fig:dist}
(b) is 0.66.
For $\epsilon=1.0$ and $\epsilon=0.3$, we plot the $I$ values as a
function of the loss rate $|\mbox{Im}(kR)|$ as shown in
Fig. \ref{fig:overlap}.
For the both $\epsilon$ values, one can confirm a clear trend that the
smaller the loss rate, the larger the overlap.

As can be seen in Fig. \ref{fig:dist} (b), $H(s,p)$ is localized
strongly on some portion of $\Omega_P$.
This localization causes the discrepancy in far-field emission
patterns between a cavity mode and the ray model.
Since the localization pattern varies depending on the mode, by
averaging $H(s,p)$ over low-loss modes, one obtains a smeared
distribution $\bar{H}(s,p)$, which agrees more closely with $P(s,p)$
than the individual $H(s,p)$.
This improvement of the ray-wave correspondence explains why the
averaged far-field emission patterns shown in Fig. \ref{fig:ffp}
correspond better with the results of the ray model.
In Fig. \ref{fig:dist} (c), we show the average of $H(s,p)$ of the 30
lowest-loss modes.
The overlap $I$ between the high-intensity regions of $\bar{H}(s,p)$
and $\Omega_P$ becomes 0.82 for $\epsilon=1.0$ and 0.73 for
$\epsilon=0.3$, which are indicated by dashed lines in
Fig. \ref{fig:overlap}.

Let us explain why it has been found for $n=1.5$ that each of the
low-loss modes always has a far-field emission pattern closely
corresponding to the results of the ray model \cite{Shinohara06,
Lebental07}.
In contrast to the case of $n=3.3$, for $n=1.5$ the high-intensity
regions of $P(s,p)$ turn out to consist of narrow stripes
\cite{Shinohara07}.
Moreover, each of the stripes is parallel to the curves of the
constant far-field angle.
As a result, irrespective of how $H(s,p)$ is localized on $\Omega_P$,
the localized regions are always located near the curves of the
constant far-field angle, and thus generates a highly directional
far-field emission pattern consistent with the result of the ray
model.
To summarize, several conditions need to be satisfied for the
correspondence between the far-field pattern of an individual low-loss
mode and the ray description.
Therefore, the ray-wave correspondence for an individual low-loss mode
on the level of far-field emission patterns can be observed only for a
specific choice of the parameters such as cavity shape and the
refractive index.
On the contrary, we expect that the agreement of the supports of
$P(s,p)$ and $H(s,p)$ can be observed more robustly for an individual
low-loss mode in a sufficiently semiclassical regime.

So far, we have shown numerically that the ray model for $n=3.3$ can
be validated when one considers the average of many low-loss cavity
modes in the semiclassical regime.
We checked that this correspondence can be also observed between
transverse electric (TE) wave calculations and the ray model with
Fresnel' law for TE polarization, the detail of which is reported
in Ref. \cite{Choi}.
Lastly, we examine whether the ray model can explain experimental
far-field data for stadium-shaped GaAs microcavity lasers when
multiple modes are involved in lasing.
We fabricated stadium-shaped microcavity lasers by using a
metalorganic chemical vapor deposition grown gradient-index,
separate-confinement-heterostructure, single-quantum-well
GaAs/Al$_x$Ga$_{1-x}$As structure and a reactive-ion-etching technique
\cite{Fukushima04}.
The radius $R$ of the stadium is fixed as 25 $\mu m$, and the
$\epsilon$ value is varied from $0.2$ to $1.0$.
Lasing is achieved at room temperature by using a pulsed current with
500-ns width at a 1 kHz repetition.
We confirmed the sharp narrowing of optical spectra around 850 nm
above the lasing threshold, and we checked that lasing occurs in
multimodes.
For instance, for the cavity with $\epsilon=1.0$ we found 13 peaks in
the spectrum.
%
%
In Fig. \ref{fig:ffp2} we plot measured far-field emission patterns.
For each $\epsilon$ value, we plot data for two different samples (red
and green curves), which are fabricated in the same way and pumped with
the same injection current strength.
We also superimpose the far-field emission patterns obtained from the
ray model in the blue curve.
In the ray model we employed Frenel's law for TE polarization, taking
into account that experimentally observed emission is TE polarized.
The resulting TE far-field patterns of the ray model are slightly
different from those of TM polarization.
In Fig. \ref{fig:ffp2}, one can confirm that changes in the
experimental far-field emission patterns due to the change in
$\epsilon$ value are well explained by the ray model.
%

In general, the relation between cavity modes and a multi-mode lasing
state formed by nonlinear mode-interaction through an active medium
can become very complicated \cite{Harayama05}.
However, the present work revealed that, at least concerning far-field
emission patterns, the average of low-loss modes approximates a
multi-mode lasing state for a sufficiently large $nkR$ value.
We believe that the enhancement of the ray-wave correspondence due to
the averaging by low-loss modes plays a significant role in yielding
the good correspondence between the experimental far-field emission
patterns and those of the ray model.

In summary, we studied light emission patterns of low-loss modes for
stadium-shaped semiconductor microcavities.
The correspondence between low-loss modes and the ray model was
revealed by investigating the phase space distributions describing
near-field emission patterns.
Close correspondence was found between experimental far-field data and
the results of the ray model, which we attributed to the enhancement
of the ray-wave correspondence due to the averaging by low-loss modes.

The work at ATR was supported in part by the National Institute of
Information and Communications Technology of Japan.
\end{document}